\documentclass[12pt]{article}
\usepackage{latexsym}
\usepackage{epsfig,amssymb,euscript,slashed}
\usepackage[linktocpage]{hyperref}
\usepackage{amsmath}
\usepackage[nosort]{cite}
\usepackage{array,calc,epsfig}
\usepackage{bbm}
\usepackage{fancybox}

\usepackage[left=1.5cm,right=1.5cm,top=3.5cm,bottom=3.5cm]{geometry}

\numberwithin{equation}{section} 
\usepackage[]{graphicx}
\usepackage{xcolor}

\begin{document}
\font\cmss=cmss10 \font\cmsss=cmss10 at 7pt

\hfill
	\vspace{18pt}
	\begin{center}
		{\Large 
			\textbf{
                         Holographic correlators with non-supersymmetric multi-particle states
                        }}
		
	\end{center}

		\vspace{8pt}
	\begin{center}
		{\textsl{Michele Giorgi}}
		\vspace{.0cm}
				
		\textit{\small Section de Math\'ematiques, Universit\'e de Gen\'eve, 1211 Gen\'eve 4, Switzerland} \\ 
		
		\vspace{.5cm}
		
		{\textsl{Stefano Giusto}}
		\vspace{.0cm}

		\textit{\small Dipartimento di Fisica,  Universit\`a di Genova, Via Dodecaneso 33, 16146, Genoa, Italy} \\  \vspace{0pt}
		
		\textit{\small I.N.F.N. Sezione di Genova,
			Via Dodecaneso 33, 16146, Genoa, Italy}\\

	\end{center}

	\vspace{12pt}

	\begin{center}
		\textbf{Abstract}
	\end{center}

We consider heavy-heavy-light-light (HHLL) holographic correlators in which the heavy states are coherent superpositions of $n$-particle operators. In the limit of large $n$ and large central charge, $c$, with $n/c$ held fixed, these HHLL correlators are described by a classical supergravity computation. When the multi-particle operators are half-BPS, it has been shown that the $n/c\to 0$ limit of the HHLL correlator encodes the tree-level connected contribution to the 4-point function with the heavy operators replaced by two $n$-particle operators with {\em finite} $n$. We provide evidence that this relation continues to hold when the multi-particle operators are non-supersymmetric. Although our checks are carried out in the AdS$_3$/CFT$_2$ setting, we expect that our conjecture extends to more general holographic dualities.

	\vspace{4pt} 
	\begin{center}
	\begin{minipage}{15.2cm}
	{\small
	\baselineskip=15pt
	\parskip=5pt
	\noindent 
				
		}
	\end{minipage}
	\end{center}	
		
		\vspace{1cm}

		\thispagestyle{empty}

		\vfill
		\vskip 5.mm
		\hrule width 5.cm
		\vskip 2.mm
		{
			\noindent {\scriptsize e-mails: michele.giorgi@unige.ch, stefano.giusto@ge.infn.it}
		}

		\setcounter{footnote}{0}
		\setcounter{page}{0}

\newpage

\section{Introduction}
Holographic dualities \cite{Maldacena:1997re,Witten:1998qj} have provided a way to compute CFT correlators in the regime of strong coupling and large central charge. For dynamical correlators with four or more external points, the holographic results are largely limited to supersymmetric operators, particularly those preserving half of the supercharges of the vacuum, that include both single-particle and multi-particle operators. 

Non-supersymmetric single-particle operators have non-protected conformal dimensions that typically diverge at strong coupling. They are dual to string states whose study requires working in the full string theory rather than in its more tractable supergravity limit. There is, however, a class of non-supersymmetric operators that can be accessed within the supergravity approximation. These are multi-particle operators whose single-particle components are gravity operators -- chiral primaries or their global descendants -- that do not preserve any common supercharges. These non-BPS multi-particle operators have an anomalous dimension that is interpreted in the gravity picture as the (negative) interaction energy between their single-particle constituents. When the number of particles, $n$, is small and fixed while the central charge, $c$, is taken to infinity, this interaction is suppressed by powers of $1/c$. In this regime, the anomalous dimensions and three-point couplings of these operators can be extracted by studying the OPE decomposition of 4-point functions containing half-BPS single or multi-particle operators \cite{Aprile:2017xsp,Aprile:2021mvq,Aprile:2026yit}. 

What we would like to compute are, instead, 4-point correlators where the non-BPS multi-particle operators appear as external states. A possible approach\footnote{For half-BPS operators in $\mathcal{N}=4$ SYM, this approach has been recently carried out in \cite{Aprile:2026uxe}, based on the 5-point functions derived in \cite{Goncalves:2019znr,Goncalves:2023oyx}.} would be to consider higher-point functions containing the single-particle constituents of the multi-particle operators evaluated at different points, and then to take the (singular) limit where some of these points collide. We suggest here a different approach, that has already been employed in the computation of correlators containing half-BPS multi-particle operators, both in AdS$_3$ \cite{Ceplak:2021wzz} and in AdS$_5$ \cite{Aprile:2024lwy,Aprile:2025hlt}. The idea is to consider the limit in which the particle number $n$ and $c$ are taken to infinity together, keeping the ratio $n/c$ fixed. In this regime the multi-particle operators become heavy, the interaction energy between the constituents becomes of order one, and the operators are described by non-trivial, asymptotically AdS geometries -- known as microstate geometries or fuzzballs \cite{Shigemori:2020yuo,Bena:2022rna} -- that reduce to global AdS when $n/c\to 0$. These geometries can sometimes be constructed exactly in the supersymmetric case\footnote{The construction of microstate geometries has been particularly developed in AdS$_3$, starting from the half-BPS case \cite{Lunin:2001jy} and extending it to quarter-BPS geometries \cite{Bena:2015bea,Bena:2016ypk,Bena:2017xbt,Heidmann:2019zws,Heidmann:2019xrd,Houppe:2020oqp,Ganchev:2021iwy,Ganchev:2022exf}; examples of exact asymptotically AdS$_5$ half-BPS geometries can be found in \cite{Lin:2004nb,Liu:2007xj,Giusto:2024trt}.} and perturbatively in $n/c$ for non-supersymmetric states \cite{Ganchev:2021ewa,Ganchev:2023sth}. The holographic two-point function of some light field in these backgrounds can be computed by solving the wave equation that describes the linearised perturbation of the background by the light field. This two-point function can also be re-interpreted as the HHLL correlator between the two light operators and the two heavy multi-particle operators \cite{Galliani:2017jlg,Bombini:2017sge}. Assume that one knows the HHLL correlator perturbatively in $n/c$. The question is if one can extrapolate this result from the regime $n\sim c\gg 1$, where it has been derived, to the regime where $n$ is finite and $c\gg 1$. If this were the case, from the HHLL correlator one could extract the LLLL correlators where the two heavy operators have been replaced by two light $n$-particle operators with finite $n$. 

Quite surprisingly, it has been found that this interpolation between different regimes produces the correct result for the correlators with half-BPS multi-particle\footnote{The same limiting procedure leads to the expected results also in the particular case in which the multi-particle operators are single-particle ones \cite{Giusto:2018ovt,Giusto:2019pxc,Turton:2024afd,Turton:2025cnn}.} states \cite{Ceplak:2021wzz, Aprile:2024lwy,Aprile:2025hlt}. More precisely, the $n/c\to 0$ limit of the HHLL correlator encodes the {\em tree-level connected} part of the correlators with light multi-particle states, associated in the bulk with connected tree Witten diagrams. At the same level in the $1/c$ expansion, the full correlators also contain non-connected loop contributions, which are not captured by a classical supergravity calculation. When the multi-particle operators are half-BPS, they can be obtained via a {\em smooth} OPE limit of $n$ single-particle operators. Hence, the 4-point function with two multi-particle operators is the limit of a correlator of $2n+2$ single-particle operators, and the standard large-$c$ counting, reviewed in the next section, can be applied to this higher-point correlator, providing a justification for the interpolation between the large-$n$ and finite-$n$ regimes. For non-supersymmetric multi-particle operators, however, the analogous OPE limit contains divergent terms, making the relation to higher single-particle correlators more subtle. This might suggest that the agreement between these distinct regimes is a consequence of supersymmetry. In this article, however, we provide evidence that the agreement extends to non-supersymmetric cases as well. 

In Section~\ref{sec:lightlimit} we present a more detailed description of the heavy states, of the associated HHLL correlators and of our conjecture relating the light limit ($n/c\to 0$) of the HHLL correlator with the LLLL correlators two containing $n$-particle operators. In Section~\ref{sec:nonBPScorr}, we test our conjecture in the non-BPS case by computing and analyzing an explicit correlator in AdS$_3$. The concluding Section~\ref{sec:concl} contains a brief summary of our findings. 

Our conjecture, if correct, enables the computation of 4-point correlators containing non-BPS multi-particle operators in a general holographic setting and provides a new approach to studying their associated CFT data, like anomalous dimensions and three-point couplings.

\section{HHLL correlators and the light limit}
\label{sec:lightlimit}

In holographic CFTs, some ``heavy" operators, $\mathcal{O}_H$, have a dual gravitational description in terms of classical geometries that asymptote AdS times some compact space. The best studied examples, and those on which we base our general picture, are the 4D $\mathcal{N}=4$ SYM theory, with central charge $c\sim N^2$, dual to AdS$_5\times$ S$^5$, and the 2D D1-D5 CFT, with $c=6N$, dual to AdS$_3\times$ S$^3\times\mathcal{M}_4$, with $\mathcal{M}_4=T^4$ or $K3$. 

Semiclassical heavy states are coherent superpositions of {\em multi-particle states} constructed from some supergravity operators $\mathcal{O}_i$ whose conformal dimensions, $\Delta_i$, are kept finite in the large $c$ limit. $\mathcal{O}_i$ could be a chiral primary operator (CPO) or one of its descendants obtained by acting on the CPO with the global generators of the chiral algebra, comprising the conformal and R-symmetry generators and the supercharges. Depending on the nature of the descendants, the multi-particle operators, $\mathcal{O}_1 \mathcal{O}_2 \ldots$, could preserve some or no supercharges. In the simplest case in which $\mathcal{O}_i$ are CPO's, the OPE between them is regular and thus one can simply define the two-particle operator, $(\mathcal{O}_1 \mathcal{O}_2)$, as
\begin{equation}
(\mathcal{O}_1 \mathcal{O}_2)(x) = \lim_{y\to x} \mathcal{O}_1(y) \mathcal{O}_2(x)\,,
\end{equation}
and analogously for $(\mathcal{O}_1\ldots \mathcal{O}_n)$ with $n>2$. When $\mathcal{O}_1$ and $\mathcal{O}_2$ are descendants, their OPE could contain singular terms. However,  if they preserve a common subset of supercharges, one can define a supersymmetric two-particle operator, $(\mathcal{O}_1 \mathcal{O}_2)$, by isolating the regular term of their OPE:
\begin{equation}
\mathcal{O}_1(y) \mathcal{O}_2(x)=(\mathcal{O}_1 \mathcal{O}_2) (x) + \dots\,,
\end{equation}
where the dots include terms that either diverge or vanish in the limit $y\to x$. Such a regular term exists because the dimension of $(\mathcal{O}_1 \mathcal{O}_2)$ is protected and equal to $\Delta_1+\Delta_2$. The situation is different for non-BPS composite operators, which have an anomalous dimension, $\Gamma_{12}$, suppressed in the large $c$ limit and thus
\begin{equation}\label{eq:nonBPSOPE}
\mathcal{O}_1(y) \mathcal{O}_2(x) = |x-y|^{\Gamma_{12}}\,(\mathcal{O}_1 \mathcal{O}_2)(x) + \ldots\quad \mathrm{with}\quad \Gamma_{12}=\frac{\gamma_{12}}{c}+O(c^{-2})\,.
\end{equation}

Consider a coherent state made with a single elementary constituent, $\mathcal{O}$, which could be either a CPO or a descendant:
\begin{equation}\label{eq:heavyOH}
\mathcal{O}_H(\alpha)=\sum_{n=0} c_n\,\alpha^n\,[\mathcal{O}^n]\,,
\end{equation}
where $\alpha$ is the coherent state parameter, which we choose to be real, $[\mathcal{O}^n]$ denotes the normalised $n$-particle operator
\begin{equation}
[\mathcal{O}^n]=\frac{(\mathcal{O}^n)}{\| (\mathcal{O}^n) \|}\,,
\end{equation}
with $(\mathcal{O}^n)$ the $n$-particle state defined as above through the OPE of $n$ operators $\mathcal{O}$, and $c_n$ are coefficients whose large $c$ limit is
\begin{equation}\label{eq:cnlargec}
c_n = \left(\frac{c^n}{n!}\right)^{\frac{1}{2}}(1+O(c^{-1}))\,.
\end{equation}
The upper limit of the sum over $n$ in \eqref{eq:heavyOH} is typically infinite for CFTs in more than two dimensions, while it is a finite number proportional to $c$ in 2D, due to the ``stringy exclusion principle" \cite{Maldacena:1998bw}. One can consider more general heavy states than the one in \eqref{eq:heavyOH}, made of several single-particle constituents, $\mathcal{O}_i$, and depending on the corresponding parameters, $\alpha_i$, and we will indeed make use of a two-parameter example in our main computation. However, we refer to the simplest one-parameter case, \eqref{eq:heavyOH}, to illustrate the main ideas.   

At large $c$ and strong coupling with fixed $\alpha$, the gravity dual of $\mathcal{O}_H(\alpha)$ is an $\alpha$-dependent, regular, asymptotically AdS geometry -- sometimes referred to as fuzzball, microstate geometry, superstratum or bubbling geometry --, that reduces at linear order in $\alpha$ to the perturbation of global AdS dual to the gravity operator $\mathcal{O}$. For finite $\alpha$, there is a unique non-linear completion of this linearised solution that does not introduce sources corresponding to single-particle operators other than $\mathcal{O}$. On the CFT side and in the same limit, the sum over $n$ in \eqref{eq:heavyOH} is peaked over an average value, $\bar n$, that is determined by $\alpha$:
\begin{equation}\label{eq:nbar}
{\bar n} = c \, f(\alpha) = c\,(\alpha^2+O(\alpha^4))\,,
\end{equation}
where $f(\alpha)$ is a function that can be deduced from the full non-linear solution, but its small $\alpha$ limit, $f(\alpha)\approx \alpha^2$, is implied by the following  simple argument. Consider the norm squared of $\mathcal{O}_H$ for $c\to \infty$
\begin{equation}
\| \mathcal{O}_H(\alpha) \|^2 \approx \sum_n \frac{(c\,\alpha^2)^n}{n!}\approx \sum_n e^{-S(n)} \quad\mathrm{with}\quad S(n)=n \log n -n -n\, \log(c\,\alpha^2)\,.
\end{equation}
The function $S(n)$ is minimized at $n={\bar n}\approx c\,\alpha^2$, in agreement with \eqref{eq:nbar}. Moreover, the spread around the average value ${\bar n}$ \begin{equation}
\Delta n \sim \left(S^{''}({\bar n})\right)^{-1/2} \approx \sqrt{\bar n}
\end{equation}
is much smaller than ${\bar n}$ when $c\gg 1$. Thus the large $c$ limit is a semi-classical limit in which the coherent state $\mathcal{O}_H$ becomes tightly peaked over an average value for the particle number, $\bar n$, of order $c$. Thus, even if $\mathcal{O}_H$ is not exactly a primary, in this limit it is well approximated by a primary with a large conformal dimension that scales like $c$: $\Delta_H\sim c$. For the same reason, one expects that the supergravity picture in which $\mathcal{O}_H$ is described by a classical geometry might provide information about multi-particle operators, $(\mathcal{O}^n)$, with a large particle number, $n\sim c$, with the tuneable parameter $\alpha$ controlling the ratio $n/c$. In the following, we will elaborate on the consequences of this picture for the holographic computation of correlators involving the multi-particle states, $(\mathcal{O}^n)$. 

As a first step, the knowledge of the gravity dual of $\mathcal{O}_H$ gives a direct way to compute HHLL correlators of the type
\begin{equation}
\mathcal{C}_H(z,{\bar z};\alpha)=\langle \mathcal{O}_H(0)\, \bar{\mathcal{O}}_H(\infty) \, \bar{\mathcal{O}}_L(1) \,\mathcal{O}_L(z,{\bar z}) \rangle\,, 
\end{equation}
where $\mathcal{O}_L$ is some single-particle gravity operator of finite dimension, $\Delta_L$; $(z,{\bar z})$ denote the two harmonic ratios on which a 4-point function depends in any d-dimensional CFT; the dependence of $\mathcal{C}_H$ on the parameter $\alpha$ is, of course, inherited from the $\alpha$ dependence of $\mathcal{O}_H$. To derive the correlator $\mathcal{C}_H$ one solves the wave equation describing the linear perturbation of the geometry of $\mathcal{O}_H$ by the gravity field dual to $\mathcal{O}_L$. The holographic recipe identifies $\mathcal{C}_H$ with the ratio between the normalizable and non-normalizable modes of the wave function expanded at the AdS boundary. As usual, the classical supergravity computation gives the correlator at leading order in the large $c$ and strong coupling expansion. 

The question we want to ask is if LLLL correlators containing two multi-particle operators, $[\mathcal{O}^n]$, with finite $n$, and two light operators, $\mathcal{O}_L$:
\begin{equation}\label{eq:Cndef}
\mathcal{C}_n(z,{\bar z}) = \langle [\mathcal{O}^n](0) \,[\bar{\mathcal{O}}^n](\infty) \,\bar{\mathcal{O}}_L(1) \,\mathcal{O}_L(z,{\bar z}) \rangle\,,
\end{equation}
are encoded in the HHLL correlator $\mathcal{C}_H$. A formal identity following from the form of the coherent state $\mathcal{O}_H$, \eqref{eq:heavyOH}, in the large $c$ limit, \eqref{eq:cnlargec},  implies that
\begin{equation}\label{eq:CHsmallalpha}
\mathcal{C}_H(z,{\bar z};\alpha) =\sum_{n=0} \frac{\alpha^{2n}}{n!}\,c^n\,\mathcal{C}_n(z,{\bar z})\,,
\end{equation}
and thus one should be able to extract $\mathcal{C}_n$ from the term of order $\alpha^{2n}$ in the small $\alpha$ expansion of $\mathcal{C}_H$. However, we saw above that the classical supergravity approximation is reliable in the limit in which both $n$ and $c$ are sent to infinity keeping the ratio $n/c$ finite; eventually, one could then take this $n/c$ ratio small and use the small $\alpha$ expansion of $\mathcal{C}_H$ in \eqref{eq:CHsmallalpha} to extract $\mathcal{C}_n$. This argument implies that the classical holographic result for the HHLL correlator does not directly encode the LLLL correlators $\mathcal{C}_n$ in the ordinary large $c$ limit in which one keeps $n$ finite while sending $c$ to infinity. 

Despite this, we will argue that the small $\alpha$ expansion of $\mathcal{C}_H$ in the supergravity approximation contains useful information on the large $c$ limit of the correlators $\mathcal{C}_n$ with finite $n$, and in particular it captures their ``connected" part, in a sense that we will make clearer shortly. The argument is neater when $(\mathcal{O}^n)$ is defined by a smooth OPE limit of $n$ coincident single-particle operators $\mathcal{O}$, for example, when $\mathcal{O}$ is a CPO, as  recalled above. In this case, the 4-point function $\mathcal{C}_n$ can be thought as the limit of the correlator with $2n+2$ operators\footnote{The factor $n!$ in the definition of $\hat{\mathcal{C}}_n$ accounts for the norm of $(\mathcal{O}^n)$, after taking the coincident limit.}
\begin{equation}
\hat{\mathcal{C}}_n=\frac{1}{n!}\, \langle \mathcal{O} \ldots \mathcal{O} \, \bar{\mathcal{O}} \ldots \bar{\mathcal{O}} \,\bar{\mathcal{O}}_L \mathcal{O}_L \rangle\,,
\end{equation}
when one takes the $n$ $\mathcal{O}$'s and the $n$ $\bar{\mathcal{O}}$'s to the same point. Thus, in this supersymmetric case, the properties of $\hat{\mathcal{C}}_n$ extend straightforwardly to $\mathcal{C}_n$. To clarify its large-$c$ behaviour, it is useful to decompose $\hat{\mathcal{C}}_n$ into its connected components: the first non-trivial terms are\footnote{We are assuming, for simplicity, that correlators that mix $\mathcal{O}_L$ with $\mathcal{O}$ and $\bar{\mathcal{O}}$ are trivial; this assumption will be satisfied in the explicit computation that we perform in this article. Moreover, as we are interested in the limit of $\hat{\mathcal{C}}_n$ in which all $\mathcal{O}$'s and all $\bar{\mathcal{O}}$'s are evaluated at the same point, we do not distinguish these points in \eqref{eq:connected}, and if we take these points to be $0$ and $\infty$, as in \eqref{eq:Cndef}, we can simply replace $\langle \mathcal{O} \bar{\mathcal{O}} \rangle \to 1$.\label{foot:ass}}
\begin{equation}\label{eq:connected}
\hat{\mathcal{C}}_n= \hat{\mathcal{C}}_n^{(c)} + n\,\langle \mathcal{O} \bar{\mathcal{O}} \rangle \,\hat{\mathcal{C}}_{n-1}^{(c)}+\frac{n(n-1)}{2}\,\langle \mathcal{O} \bar{\mathcal{O}} \rangle^2 \,\hat{\mathcal{C}}_{n-2}^{(c)}+\ldots\,,
\end{equation} 
where the subscript $(c)$ denotes a fully connected correlator, according to the usual quantum field theory definition. In the large $c$ expansion, $\hat{\mathcal{C}}_n^{(c)}$ starts at order $c^{-n}$, corresponding to the supergravity tree-level term, while every loop correction is further suppressed by $c^{-1}$. Since in the small $\alpha$ expansion of $\mathcal{C}_H$, \eqref{eq:CHsmallalpha}, the correlator $\mathcal{C}_n$ is multiplied by $c^n$, only the terms of order $c^{-n}$ of $\mathcal{C}_n$ might contribute to the HHLL correlator. Terms like the tree-level parts of $\hat{\mathcal{C}}_{n-1}^{(c)}$, the tree-level and one-loop parts of $\hat{\mathcal{C}}_{n-2}^{(c)}$ and so on, which would give divergent contributions to $\mathcal{C}_H$, must not be included in the supergravity computation. On the other hand, there are various terms of order $c^{-n}$ on the r.h.s. of \eqref{eq:connected}: the tree-level part of $\hat{\mathcal{C}}_n^{(c)}$ but also the 1-loop part of $\hat{\mathcal{C}}_{n-1}^{(c)}$, the 2-loop part of $\hat{\mathcal{C}}_{n-2}^{(c)}$, and so on. In the standard large $c$ expansion with fixed $n$, all these terms should be included in $\mathcal{C}_n$. However, it is clear that the classical supergravity computation cannot encode the loop corrections, so the only contribution to $\mathcal{C}_n$ that can be captured by the classical HHLL correlator is the tree-level connected component, $\hat{\mathcal{C}}_n^{(c)}$. Consistency with our previous observation -- that the supergravity picture of the heavy operator, $\mathcal{O}_H$, is appropriate in the regime in which $c$ and $n$ are taken to infinity at the same rate -- requires that the large $c$ limit of $\hat{\mathcal{C}}_n^{(c)}$ with fixed $n$ coincides with the limit with fixed $n/c$, while this is not true for the full correlator $\hat{\mathcal{C}}_n$.

To summarise, in the small $\alpha$ expansion of the supergravity HHLL correlator, only the connected part, $\mathcal{C}_n^{(c)}$, of the correlator in \eqref{eq:Cndef} must appear, and thus the identity \eqref{eq:CHsmallalpha} cannot represent the classical supergravity result. We obtain a consistent identity if we assume that what supergravity computes is actually the normalised connected two-point function of $\mathcal{O}_L$ and $\bar{\mathcal{O}}_L$ in the $\mathcal{O}_H$ background:
\begin{equation}\label{eq:CHsugra}
\mathcal{C}_H^\mathrm{sugra}(z,{\bar z};\alpha)\equiv \frac{\langle \mathcal{O}_H(0)\, \bar{\mathcal{O}}_H(\infty) \,\bar{\mathcal{O}}_L(1) \,\mathcal{O}_L(z,{\bar z}) \rangle}{\langle \mathcal{O}_H \bar{\mathcal{O}}_H \rangle}-\frac{\langle \mathcal{O}_H(0)\, \bar{\mathcal{O}}_H(\infty) \,\bar{\mathcal{O}}_L(1) \rangle\,\langle \mathcal{O}_H(0)\, \bar{\mathcal{O}}_H(\infty) \,\mathcal{O}_L(z,{\bar z}) \rangle}{\langle \mathcal{O}_H \bar{\mathcal{O}}_H \rangle^2}\,,  
\end{equation}
where the second term on the r.h.s. subtracts the one-point functions of $\mathcal{O}_L$ and $\bar{\mathcal{O}}_L$; in the simplifying assumption of footnote \ref{foot:ass}, these one-point functions vanish and thus we omit them in the following. Using again the small $\alpha$ and large $c$ expansion for $\mathcal{O}_H$, \eqref{eq:heavyOH} and \eqref{eq:cnlargec}, one obtains
\begin{equation}\label{eq:CHsugrasmallalpha}
\mathcal{C}_H^\mathrm{sugra}(z,{\bar z};\alpha)=\sum_{n=0}\frac{\alpha^{2n}}{n!}\,c^n\,\mathcal{C}_n^{(c)}(z,{\bar z})\,,
\end{equation}
where the connected correlator $\mathcal{C}_n^{(c)}$ is defined by taking the coincident OPE limit of \eqref{eq:connected}. The above identity could be seen as a formal derivation of the expectation that the classical HHLL correlator encodes the tree-level connected parts of the correlators with two $n$-particle and two single-particle states, at least when the single-particle constituent, $\mathcal{O}$, is a CPO. This expectation has been verified by several explicit calculations, both in AdS$_3$ \cite{Ceplak:2021wzz} and in AdS$_5$ \cite{Aprile:2024lwy,Aprile:2025hlt}.

\subsection{Correlators with non-BPS multiparticle operators}

Whether the identity \eqref{eq:CHsugrasmallalpha} applies also when $\mathcal{O}$ is a descendant of a CPO, and especially in cases in which the multi-particle states, $(\mathcal{O}^n)$, do not preserve any supersymmetry, is still unverified and far from obvious; in particular, subtleties might arise from the singular OPE limit that is implied in the definition of the multi-particle operators (see \eqref{eq:nonBPSOPE}). Our purpose in this article is to test \eqref{eq:CHsugrasmallalpha} for some non-BPS multi-particle operator: we will compute in gravity the HHLL correlator in the small $\alpha$ expansion, assume the identity \eqref{eq:CHsugrasmallalpha} to formulate a prediction for the tree-level connected part of a correlator containing two non-BPS multi-particle operators and two light gravity operators and perform a consistency check on this prediction. When dealing with correlators at strong coupling with non-supersymmetric  external states, there are not many quantitative tests one can perform. We will verify a basic property of the correlator that, despite its simplicity, we believe to be quite non-trivial. We sketch here the idea of the check, and provide the details after we present the explicit computation.  

The leading term in the $\bar z\to 1$ expansion of the correlator $\mathcal{C}_n^{(c)}(z,{\bar z})$ with $z$ finite is given by the conformal block of the identity. Hence, it is completely determined by conformal invariance in terms of the dimensions of the external operators, and in particular of the dimension of the non-BPS multi-particle operator, $(\mathcal{O}^n)$, which, as we saw, contains an anomalous contribution:
\begin{equation}
\Delta_{(\mathcal{O}^n)}= n\,\Delta_\mathcal{O}+ \frac{\gamma_n}{c}+O(c^{-2})\,.
\end{equation}
Thus, by studying the ${\bar z}\to 1$ limit of the correlator derived from gravity, one can extract the anomalous dimension, $\gamma_n$. If one has an independent way to compute $\gamma_n$, one can either disprove or provide evidence on the consistency of the gravity result. For example, one can exploit the fact that double-particle operators can be exchanged in the OPE of two single-particle operators to deduce their anomalous dimension from the 4-point correlators containing those single-particles as external states. We will consider in the following section a double-particle operator whose anomalous dimension can be easily computed with this method from a known correlator with single-particle operators in AdS$_3$. We will compare this anomalous dimension with the one deduced from the correlator in which the double-particle operators appear as external states, computed using \eqref{eq:CHsugrasmallalpha}, and find agreement. Note that, in the non-supersymmetric case, the heavy state $\mathcal{O}_H$ has itself an anomalous dimension:
\begin{equation}\label{eq:heavydim}
\Delta_H = {\bar n}\, \Delta_\mathcal{O} + \frac{{\bar n}^2}{c}\,\gamma^{(1)}_H + \frac{{\bar n}^3}{c^2}\,\gamma^{(2)}_H+\ldots
\end{equation}
where the corrections to the bare value ${\bar n}\, \Delta_\mathcal{O}$, coming from the interaction between the $\bar n$ elementary constituents $\mathcal{O}$, are proportional to ${\bar n}/c \approx \alpha^2$, and are thus finite in the supergravity regime in which ${\bar n}\sim c$. The anomalous dimensions $\gamma_H^{(i)}$ are naturally encoded in the geometry dual to $\mathcal{O}_H$, and can be derived by computing its ADM mass \cite{Ganchev:2021ewa,Ganchev:2023sth}. The anomalous dimensions $\gamma_n$ of the ``light" multi-particle operators $(\mathcal{O}^n)$, refer instead to a very different regime, in which $n$ is kept finite as $c$ is sent to infinity. Hence, a priori one does not expect a relation between $\gamma_n$ and $\gamma^{(i)}_H$. However, by working out a particular example, we will see that the naive extrapolation of \eqref{eq:heavydim} to finite $\bar{n}$ yields the correct anomalous dimension $\gamma_n$.  

\section{A correlator with non-BPS bound states in AdS$_3$}
\label{sec:nonBPScorr}

We present here the computation of a correlator containing non-BPS multi-particle operators, using \eqref{eq:CHsugrasmallalpha}, and show that it passes the test discussed at the end of the previous section. The gravity derivation of the HHLL correlator works very similarly to the supersymmetric cases that have been considered both in AdS$_3$ \cite{Bombini:2017sge,Ceplak:2021wzz} and in AdS$_5$ \cite{Aprile:2025hlt}. We focus on the AdS$_3$ setting, where the geometries dual to some non-supersymmetric heavy states have already been constructed \cite{Ganchev:2021pgs,Ganchev:2021ewa,Ganchev:2023sth}, exploiting the fact that the lowest dimensional CPO's and their descendants can be described within a tractable 3D consistent truncation \cite{Mayerson:2020tcl}. The 6D supergravity compactified on $T^4$ or $K3$  contains $N_f$ tensor multiplets, with $N_f=5$ for $T^4$ and $N_f=21$ for $K3$; the lowest harmonics of these fields on S$^3$ give rise to operators of left-right dimension $(h,\bar{h})=(1/2,1/2)$, which we denote by $O_f^{\alpha \dot{\alpha}}$, with $f=1,\ldots,N_f$ being the flavour index, and $(\alpha,\dot{\alpha})=(\pm,\pm)$ being indices in the fundamental representation of the R-symmetry group $SU(2)_L\times SU(2)_R$. Thus $O_f^{++}$ is a CPO of dimension $\Delta=h+\bar{h}=1$. 

To construct non-supersymmetric multi-particle states, one could consider, for example, the descendant $\mathcal{O}=L_{-1} {\tilde L}_{-1}O_f^{++}$, obtained by acting with the left and right moving Virasoro generators of the global $SL(2,\mathbb{C})$ symmetry. The single-particle $\mathcal{O}$ is still a protected operator, but the multi-particle operators, $(\mathcal{O})^n$, constructed with it are non-BPS: in the gravity picture, the $\mathcal{O}$ particles have more mass than charge and attract each other. The geometry dual to the heavy state made out of this $\mathcal{O}$ has been constructed in \cite{Ganchev:2021ewa}, and one can solve the wave equation in this background and extract the correlators $\mathcal{C}_n^{(c)}$ using \eqref{eq:CHsugrasmallalpha}. We have performed this computation for $n=2$ and we provide some details in Appendix~\ref{sec:appA} and the result in an ancillary file. However, $(\mathcal{O})^2$ has a quite large bare dimension, equal to 6, and computing its anomalous dimension\footnote{More precisely, $(\mathcal{O})^2$ is not a primary, but the superposition of some primaries and some descendants; each of these have their own anomalous dimensions.}, so as to perform the test described above, is complicated by the mixing with other operators with the same bare dimension.  Thus, we leave this task for the future. 

To simplify the computation of the anomalous dimension, it is useful to consider a double-particle operator with a smaller bare dimension. We could consider, for example, the two descendants
 \begin{equation}
 \mathcal{O}_1 = L_{-1} O^{++}_f\quad, \quad \mathcal{O}_2 = O_f^{+-}\,.
 \end{equation}
 They do not preserve any common supersymmetry ($\mathcal{O}_1$ breaks the left and $\mathcal{O}_2$ the right supersymmetries) and thus the double-particle $(\mathcal{O}_1 \mathcal{O}_2)$ is non-BPS and it has bare dimension equal to 3. As we will see, it is immediate to extract its anomalous dimension from a known 4-point function containing the single-particle operators $\mathcal{O}_1$ and $\mathcal{O}_2$. The gravity derivation of the correlator with this choice of double-particle and the test of this result are described in the following subsections.  

\subsection{The correlator from gravity}
To describe the double-particle $(\mathcal{O}_1 \mathcal{O}_2)$ one needs a slight generalization of the coherent state $\mathcal{O}_H$, depending on two parameters $\alpha_1$ and $\alpha_2$:
\begin{equation}\label{eq:OHO1O2}
\mathcal{O}_H = 1+N^{\frac{1}{2}} \,\alpha_1\,\mathcal{O}_1 + N^{\frac{1}{2}}\,  \alpha_2\,\mathcal{O}_2 + N \alpha_1 \alpha_2 \,[\mathcal{O}_1\mathcal{O}_2]+\ldots
\end{equation}
where one should remember that in the D1-D5 CFT $c=6 N$; the operators $\mathcal{O}_1$, $\mathcal{O}_2$ and $[\mathcal{O}_1\mathcal{O}_2]$ have unit norm. 

The geometry dual to this state has been constructed in \cite{Ganchev:2023sth} (where it corresponds to the case $n_1=1$, $n_2=0$ of Section 3.2.2). We summarize here the relevant features of the construction. The solution sits in a 3D consistent truncation that includes an asymptotically AdS$_3$ metric:
\begin{equation}
ds^2_3 = -\Omega_1^2 \left(d\tau+\frac{k}{1-\xi^2} d\psi\right)^2+\frac{\Omega_0^2}{(1-\xi^2)^2}(d\xi^2+\xi^2 d\psi^2)\,,
\end{equation}
and, in a particular gauge, eight scalars, denoted as $\nu_i$, $\mu_i$, $\lambda_i$ ($i=1,2$), $m_5$, $m_6$, and four vectors $A^{12}$, $A^{34}$, $A^{23}$, $A^{14}$ (with $A^{IJ}_\xi=0$). The crucial simplifying feature of the gravity ansatz (also known as ``Q-ball ansatz") is that all the dynamical fields can be taken to depend only on the radial coordinate $\xi\in[0,1]$. The dependence on the AdS$_3$ time, $\tau$, and angular variable, $\psi\in[0,2\pi]$, is fixed by phases determined by the bare dimensions of the fields $\mathcal{O}_1$ and $\mathcal{O}_2$; moreover, these phases can be re-absorbed in a constant shift of the gauge fields $A^{12}$ and $A^{34}$. Thus the supergravity equations reduce to a system of coupled ordinary differential equations, which can be solved perturbatively in $\alpha_1$, $\alpha_2$ (or numerically). Not all non-BPS heavy states are described by the Q-ball ansatz, but the state $O_H$ with our choice of $\mathcal{O}_1$ and $\mathcal{O}_2$, can. The most important fields for our purposes are the 3D metric, which will determine the wave equation, and the scalars $\nu_1$ and $\nu_2$, which describe the fields dual to $\mathcal{O}_1$ and $\mathcal{O}_2$. When $\alpha_i=0$ the solution must reduce to global AdS$_3$, which corresponds to
\begin{equation}\label{eq:ds30}
\Omega_0=\Omega_1=1\quad,\quad k=\xi^2\,,
\end{equation}
after redefining the angular variable as $\theta = \psi-\tau$. At linear order in $\alpha_i$, one has a perturbation of AdS given by 
\begin{equation}\label{eq:nu1}
\nu_1 = \tilde{\alpha}_1\,\xi\quad,\quad \nu_2 = \tilde{\alpha}_2\,.
\end{equation}
We denote the gravity parameters by $\tilde{\alpha}_i$ to distinguish them from the CFT parameters $\alpha_i$ that appear in \eqref{eq:OHO1O2}; at linear order, the two sets of parameters are proportional but they might be related by a non-trivial redefinition at higher orders. Determining the precise map between  $\tilde{\alpha}_i$ and $\alpha_i$ will be crucial for our check and we will come back to this point at the end of the section.  Starting from \eqref{eq:ds30} and \eqref{eq:nu1}, one can solve the equations of motion order by order in $\tilde{\alpha}_i$ requiring the solution to be regular everywhere and to decay in a normalizable way at the AdS boundary, $\xi=1$. There is a unique solution with these properties, up to redefinitions of $\tilde{\alpha}_i$. 

Our goal is to compute the correlator with two insertions of $(\mathcal{O}_1 \mathcal{O}_2)$ and two insertions of some light operator $\mathcal{O}_L$, which we would like to choose to make the computation as simple as possible. One can consider the scalar of vanishing R-charge and dimension $(1,1)$ obtained by acting with the supercharges on a $(1/2,1/2)$ CPO $O^{++}_g$ with a flavour $g$ {\it different} from $f$ (the flavour corresponding to $\mathcal{O}_I$): 
\begin{equation}\label{eq:OLdef}
\mathcal{O}_L = G^{-1} \tilde{G}^{-1} O^{++}_g\,.
\end{equation}
In the gravity picture, $\mathcal{O}_L$ is described by a minimally coupled scalar, $\Phi_L$, in the 6D geometry of $\mathcal{O}_H$ \cite{Bombini:2017sge} and, since we are considering the lowest S$^3$ harmonic,\footnote{Correlators involving higher harmonics could also be computed by considering the 6D version of the wave equation \eqref{eq:wave3D}.} its linearised equation is simply 
\begin{equation}\label{eq:wave3D}
\Box_3 \Phi_L=0\,,
\end{equation}
with $\Box_3$ the d'Alembertian computed with the metric $ds^2_3$. Since we want two insertions of $(\mathcal{O}_1 \mathcal{O}_2)$, we need this 3D metric up to order $\tilde{\alpha}_1^2 \tilde{\alpha}_2^2$. One finds
\begin{equation}
\begin{aligned}
\Omega_0 &= 1 -\frac 18\left(1-\xi^2\right)\left(\tilde{\alpha}_1^2\xi^2 + \tilde{\alpha}_2^2+\frac{\tilde\alpha_1^2\,\tilde\alpha_2^2}{72}\left(17+\xi^4\right)\right)\,,\\
\Omega_1 &= 1 - \frac 14\left(\tilde\alpha_1^2+ \tilde\alpha_2^2+\frac{\tilde\alpha_1^2\,\tilde\alpha_2^2}{72}\left(3+8\xi^2-\xi^4\right)\right) \,,\\
k &= \xi^2 + \frac{\xi^2}{4} \left(\tilde\alpha_1^2+ \tilde\alpha_2^2+\frac{\tilde\alpha_1^2\,\tilde\alpha_2^2}{36}\left(21-\xi^2+3\xi^4\right)\right)\,.
\end{aligned}
\end{equation}

Exploiting the invariance of the metric under shifts of $\tau$ and $\theta$, one can turn \eqref{eq:wave3D} into an ODE for the function $\phi(\xi)$:
\begin{equation}
\Phi_L = \sum_{l=-\infty}^\infty \int_{-\infty}^\infty \frac{d\omega}{2\pi} \,e^{i \omega \tau + i l \theta}\,\phi(\xi)\,,
\end{equation}
where $\phi(\xi)$ satisfies
\begin{equation}\label{eq:wephi}
\frac{(1-\xi^2)^2}{\xi \,\Omega_0^2 \,\Omega_1}\partial_\xi(\xi \,\Omega_1 \,\partial_\xi \phi)+\left(\frac{(\omega-l)^2}{\Omega_1^2} - \frac{(k \,\omega -(1+k-\xi^2)\,l)^2}{\xi^2 \, \Omega_0^2}\right) \phi=0\,.
\end{equation}
The smoothness of the background metric, $ds^2_3$, implies the wave function to be regular too; so one should select the solution of \eqref{eq:wephi} such that $\phi(\xi) \to \xi^{|l|}$ for $\xi\to 0$. The HHLL correlator is encoded in the limit of this solution at the AdS boundary, $\xi\to 1$. To study this limit it is useful to introduce the canonical Fefferman-Graham coordinate, $w$, in terms of which the boundary metric at $w\to 0$ becomes 
\begin{equation}
ds^2_3= \frac{dw^2}{w^2}+\frac{-d\tau^2+d\theta^2}{w^2} + O(w^0)\,.
\end{equation}
The relation between the coordinates $\xi$ and $w$ is (up to order $\tilde\alpha_1^2 \,\tilde\alpha_2^2$)
\begin{equation}\label{eq:FG}
\xi^2 = 1- w^2\left(1-\frac{1}{4}\left(\tilde\alpha_1^2 + \tilde\alpha_2^2\right)-\frac{5}{144}\tilde\alpha_1^2 \,\tilde\alpha_2^2 \right)  +O(w^4)\,.
\end{equation}
Close to the boundary $\phi$ has a ``source" term, proportional to $w^0$, and a ``VEV" term, which goes like $w^2$, as appropriate for an operator $\mathcal{O}_L$ of dimension 2:
\begin{equation}\label{eq:asphi}
\phi \underset{w\to 0}{\rightarrow} A(\omega,l) + B(\omega,l)\,w^2\,.
\end{equation}
The HHLL correlator is given by the ratio between VEV and source, $B/A$. Going back to configuration space, one thus would like to compute
\begin{equation}\label{eq:csugraFourier}
\mathcal{C}_H^\mathrm{sugra}(z,\bar{z};\tilde{\alpha}_1,\tilde{\alpha_2}) = \mathcal{N}\,(z\bar{z})^{-1} \sum_{l=-\infty}^\infty \int_{-\infty}^\infty \frac{d\omega}{2\pi} \,z^{\frac{\omega+l}{2}}\,\bar{z}^{\frac{\omega-l}{2}}\frac{B(\omega,l)}{A(\omega,l)}\,,
\end{equation}
where we have introduced the standard coordinates on the complex plane
\begin{equation}
z= e^{i(\tau+\theta)}\quad,\quad \bar{z}= e^{i(\tau-\theta)}\,,
\end{equation}
and the factor $(z\bar{z})^{-1}$ is the Jacobian relating the correlator on the cylinder to the one on the plane; $\mathcal{N}$ is a normalization constant independent of  $\tilde{\alpha}_i$ that will be fixed by requiring that $\mathcal{C}_H^\mathrm{sugra} = |1-z|^{-4}$ for $\tilde{\alpha}_i=0$.  Computing the coefficients $A$ and $B$ is mathematically equivalent to solving the connection problem between the points $\xi=0$ and $\xi=1$ for the ODE \eqref{eq:wephi}. At every order in the small $\tilde{\alpha}_i$ expansion, this equation can be mapped to a hypergeometric equation, for which the solution of the connection problem is well-known. To achieve this goal one has to perform both a change of coordinate $\xi \to x$ and a rescaling of the wave function $\phi\to \hat{\phi}$:
\begin{equation}\label{eq:changehyper}
\xi= x^{\frac{1}{2}} \,(1+(1-x)f_1(x)) \quad,\quad \phi(\xi) = x^{\frac{|l|}{2}} (1+ f_2(x))\,\hat{\phi}(x)\,,
\end{equation}
where both $f_1(x)$ and $f_2(x)$ are polynomials in $x$ at every order in $\tilde{\alpha}_i$ and vanish for $\tilde{\alpha}_i=0$. After this change of variables one finds
\begin{equation}
\hat{\phi}(x)={}_2F_1\left(\frac{|l|+\omega}{2}+a,\frac{|l|-\omega}{2} -a,1+|l| ; x\right)\,,
\end{equation}
with $a$ a function of $\omega$, $l$ and $\tilde{\alpha}_i$ that vanishes for $\tilde{\alpha}_i=0$. To evaluate the boundary limit $x\to 1$, one can use the relation
\begin{equation}\label{eq:ashyper}
{}_2F_1(\hat{a},\hat{b},\hat{a}+\hat{b}+1;x)\approx\frac{\Gamma(\hat{a}+\hat{b}+1)}{\Gamma(\hat{a}+1)\Gamma(\hat{b}+1)}\left[1+\hat{a}\,\hat{b}\, (1-x)\, \left(H_{\hat{a}}+H_{\hat{b}}-1+\log(1-x)\right)\right]\,,
\end{equation}
where
\begin{equation}
H_{\hat{a}}=\sum_{k=1}^\infty\left(\frac{1}{k}-\frac{1}{k+\hat{a}}\right)\,.
\end{equation}
From \eqref{eq:ashyper} one can immediately read off the coefficients $A$ and $B$ of \eqref{eq:asphi}:
\begin{equation}\label{eq:AoverB}
\frac{B}{A} = -\frac{1-\frac{1}{4}\left(\tilde\alpha_1^2+\tilde\alpha_2^2\right)-\frac{5}{144}\tilde\alpha_1^2\tilde\alpha_2^2}{1-2 f_1(1)}\left[\frac{l^2}{4} - \left(\frac{\omega}{2}+a\right)^2 \right]\sum_{k=1}^\infty\left(\frac{1}{k+\frac{|l|+\omega}{2}+a} +\frac{1}{k+\frac{|l|-\omega}{2}-a}\right)\,,
\end{equation}
where we have discarded terms that give only contact term contributions to the correlator $\mathcal{C}_H^\mathrm{sugra}$. Note that the factor in front of the square parenthesis comes from the relation between $(1-x)$ and $w$ (using \eqref{eq:FG} and \eqref{eq:changehyper}); this factor leads to the quite remarkable simplification:
\begin{equation}
\frac{1-\frac{1}{4}\left(\tilde\alpha_1^2+\tilde\alpha_2^2\right)-\frac{5}{144}\tilde\alpha_1^2\tilde\alpha_2^2}{1-2 f_1(1)}\left[\frac{l^2}{4} - \left(\frac{\omega}{2}+a\right)^2 \right] = \frac{l^2-\omega^2}{4}\,,
\end{equation}
that one can verify order by order in $\tilde{\alpha}_i$. When inserted in \eqref{eq:csugraFourier}, the factor $\frac{l^2-\omega^2}{4}$ turns into the differential operator $\partial_z \partial_{\bar z}$. Then one finds
\begin{equation}
\mathcal{C}_H^\mathrm{sugra}(z,\bar{z};\tilde{\alpha}_1,\tilde{\alpha}_2) = \mathcal{N}\,\partial_z \partial_{\bar z}\,\sum_{k=1}^\infty \sum_{l=-\infty}^\infty \int_{-\infty}^\infty \frac{d\omega}{2\pi} \,z^{\frac{\omega+l}{2}}\,\bar{z}^{\frac{\omega-l}{2}}\,\left(\frac{1}{k+\frac{|l|+\omega}{2}+a} +\frac{1}{k+\frac{|l|-\omega}{2}-a}\right)\,.
\end{equation}
The integral over $\omega$ can be performed with the Cauchy theorem: for $\tau>0$, we pick the poles, $\omega_k$, on the negative real axis:
\begin{equation}
k+\frac{|l|+\omega}{2}+a \,\Bigr|_{\omega=\omega_k}=0\,.
\end{equation}
The sums over $k$ and $l$ can be performed with the method of \cite{Ceplak:2021wzz}: they produce polylogarithms of $z$ and $\bar z$ of order increasing with the powers of $\tilde{\alpha}_i$. Exactly as for the supersymmetric correlators, all the polylogs arrange to give the ladder integrals \cite{Isaev:2003tk}, $\mathcal{P}_L(z,\bar{z})$,  and their derivatives. For our purposes, we will need only the first two integrals:
\begin{equation}\label{eq:ladder}
\begin{aligned}
{\cal P}_{1}(z,\bar{z})&= \log(z\bar{z}) \big( \log(1-z) -\log(1-\bar{z}) ) + 2 \big( {\rm Li}_{2}(z) - {\rm Li}_2(\bar{z})\big) \,,\\
{\cal P}_{2}(z,\bar{z})&=\frac{1}{2}\log^2(z\bar{z}) \big( {\rm Li}_{2}(z) - {\rm Li}_2(\bar{z})\big) - 3 \log(z\bar{z})\big( {\rm Li}_{3}(z) - {\rm Li}_3(\bar{z})\big) + 6 \big( {\rm Li}_{4}(z) - {\rm Li}_4(\bar{z})\big)\,.
\end{aligned}
\end{equation}  
Expanding $\mathcal{C}_H^\mathrm{sugra}(z,\bar{z};\tilde{\alpha}_1,\tilde{\alpha_2})$ in powers of $\tilde{\alpha}_i$
\begin{equation}\label{eq:Cnm}
\mathcal{C}_H^\mathrm{sugra}(z,\bar{z};\tilde{\alpha}_1,\tilde{\alpha_2}) =\sum_{n,m=0} \frac{\tilde{\alpha}_1^{2n}\, \tilde{\alpha}_2^{2m}}{n!\,m!}\,N^{n+m}\,\tilde{C}_{n,m}(z,\bar{z})\,,
\end{equation}
defines the functions $\tilde{C}_{n,m}(z,\bar{z})$; we choose the normalization such that $\tilde{C}_{0,0}=|1-z|^{-4}$. To relate these functions to the correlators with multi-particle operators, we need the map between the gravity parameters $\tilde{\alpha}_i$ and CFT parameters $\alpha_i$ that define the coherent state $\mathcal{O}_H$ in \eqref{eq:OHO1O2}. This map, at the order relevant for us, is given by eqs. (5.14) and (5.16) of \cite{Ganchev:2023sth},\footnote{The translation of symbols between \cite{Ganchev:2023sth} and this paper is the following: $\alpha_i^\mathrm{there}\to \tilde{\alpha}_i^\mathrm{here}$, $\left(\frac{N_i}{N}\right)^\mathrm{there}\to (\alpha_i^2)^\mathrm{here}$ .} which were obtained by matching the momentum and energy of $\mathcal{O}_H$ between gravity and the CFT. Note however that eq. (5.16) does not apply to $n_1=0$, and thus one has to adapt the calculation of \cite{Ganchev:2023sth} to our case ($n_1=1,n_2=0$). We find:
\begin{equation}\label{eq:mapaat}
\tilde{\alpha}_1= 2 \alpha_1\left(1-\frac{5}{36}\,\alpha_2^2\right)\quad,\quad \tilde{\alpha}_2=2 \alpha_2\left( 1+\frac{1}{36}\,\alpha_1^2\right)\,.
\end{equation}
The connected part of the correlator with two insertions of $[\mathcal{O}_1\mathcal{O}_2]$, which we will denote as $\mathcal{C}^{(c)}_{[12]}$, is given, according to \eqref{eq:CHsugrasmallalpha}, by the term $\alpha_1^2\alpha_2^2$ of $\mathcal{C}_H^\mathrm{sugra}$. Using \eqref{eq:Cnm} and \eqref{eq:mapaat}, this yields
\begin{equation}\label{eq:Cc12}
\mathcal{C}^{(c)}_{[12]}=16\left(\tilde{C}_{1,1}-\frac{5}{72}\,\frac{\tilde{C}_{1,0}}{N} + \frac{1}{72}\,\frac{\tilde{C}_{0,1}}{N}\right)\,,
\end{equation}
which we can write in terms of the ladder integrals \eqref{eq:ladder} times rational functions of $(z,\bar{z})$, $R_i$:
 \begin{equation}\label{eq:C12conn}
 \begin{aligned}
  \mathcal{C}^{(c)}_{[12]} &= \partial_z \partial_{\bar{z}} \Bigl [R_1\, {\cal P}_{2} +R_2  (z\partial_z-\bar{z}\partial_{\bar{z}}) {\cal P}_{2}+R_3 \, {\cal P}_{1}\log|z|^2+ R_4 \log^2|z|^2+R_5 \log|z|^2 \log|1-z|^2\\
 &+R_6\, {\cal P}_{1} + R_7 \log|z|^2 + R_8 \log|1-z|^2+R_9\Bigr]\,.
 \end{aligned}
 \end{equation}
We provide the functions $R_i$ in the attached ancillary file.

\subsection{Check on the anomalous dimension}
\label{sec:ADcheck}
 The gravity computation described above leads to an explicit prediction, \eqref{eq:C12conn}, for the connected correlator with two insertions of the non-BPS double-particle $ [\mathcal{O}_1 \mathcal{O}_2]$, defined as
 \begin{equation}\label{eq:Cc12b}
 \begin{aligned}
 \mathcal{C}^{(c)}_{[12]}(z,\bar{z}) &= \langle [\mathcal{O}_1 \mathcal{O}_2](0) [\bar{\mathcal{O}}_1 \bar{\mathcal{O}}_2](\infty) \bar{\mathcal{O}}_L(1) \mathcal{O}_L(z,\bar{z})\rangle\\
 &- \langle \mathcal{O}_1(0) \bar{\mathcal{O}}_1 (\infty) \bar{\mathcal{O}}_L(1) \mathcal{O}_L(z,\bar{z})\rangle- \langle \mathcal{O}_2(0) \bar{\mathcal{O}}_2(\infty) \bar{\mathcal{O}}_L(1) \mathcal{O}_L(z,\bar{z})\rangle+ \langle \bar{\mathcal{O}}_L(1) \mathcal{O}_L(z,\bar{z})\rangle\,.
 \end{aligned}
 \end{equation}
 We will perform a check on this correlator by following the logic of Section 4.1.1 of \cite{Ceplak:2021wzz}. The idea is that the limit of the correlator for $\bar{z}\to 1$ with fixed $z$ is captured, at leading order, by the Virasoro descendants of the identity, and thus it can be expressed in terms of the known vacuum Virasoro block \cite{Fitzpatrick:2015qma} and the conformal dimensions of the external operators. 
 
 However, to apply this logic one needs the external operators to be Virasoro primaries, and this is not the case neither for $\mathcal{O}_1 = L_{-1} O^{++}_f$, nor for our double-particle, $[\mathcal{O}_1\mathcal{O}_2]$, which can be decomposed into the sum of a primary, $[\mathcal{O}_1\mathcal{O}_2]_P$, and a descendant:
 \begin{equation}\label{eq:primdef}
 [\mathcal{O}_1\mathcal{O}_2]= \frac{1}{\sqrt{2}} \left([\mathcal{O}_1\mathcal{O}_2]_P + \frac{1}{\sqrt{2}}\,L_{-1}(O^{++}_f O^{+-}_f)\right)\quad \mathrm{with}\quad [\mathcal{O}_1\mathcal{O}_2]_P =\frac{1}{\sqrt{2}}\left( L_{-1}O^{++}_f O^{+-}_f-O^{++}_f L_{-1} O^{+-}_f\right)\,.
 \end{equation}
 Moreover one can compute the correlators with descendants via the Ward identity associated to $L_{-1}$:
 \begin{equation}
 \langle L_{-1} O_h(0) L_{-1} \bar{O}_h (\infty) \bar{\mathcal{O}}_L(1)\mathcal{O}_L(z,\bar{z})\rangle = \mathcal{D}_h\left[ \langle O_h(0)  \bar{O}_h (\infty) \bar{\mathcal{O}}_L(1)\mathcal{O}_L(z,\bar{z})\rangle\right]\,,
  \end{equation}
  with 
   \begin{equation}
   \mathcal{D}_h  \cdot= (1-z)^{2-2h_L} \,\partial_z ( z\, \partial_z\, ( (1-z)^{2h_L}\,\cdot) ) + 2 h\, \cdot\,,
  \end{equation}
 where $h$ and $h_L$ are the dimensions of the primaries $O_h$ and $\mathcal{O}_L$.  To simplify the notation, from now on we will omit to write the worldsheet points in the 4-point functions, assuming the standard order $(0,\infty, 1, z)$. Then the first line in \eqref{eq:Cc12b} can be decomposed as
  \begin{equation}\label{eq:primdesc}
  \begin{aligned}
  \langle [\mathcal{O}_1 \mathcal{O}_2] [\bar{\mathcal{O}}_1 \bar{\mathcal{O}}_2] \bar{\mathcal{O}}_L \mathcal{O}_L\rangle&=\frac{1}{2} \langle [\mathcal{O}_1 \mathcal{O}_2]_P\,[\bar{\mathcal{O}}_1 \bar{\mathcal{O}}_2]_P \,\bar{\mathcal{O}}_L \mathcal{O}_L\rangle+ \frac{1}{4}\mathcal{D}_1  \langle [O_f^{++} O_f^{+-}]  [O_f^{--} O_f^{-+}]  \,\bar{\mathcal{O}}_L \mathcal{O}_L\rangle\\
  &+\frac{1}{2\sqrt{2}} \langle [\mathcal{O}_1 \mathcal{O}_2]_P\, L_{-1}[O_f^{--}O_f^{-+}] \,\bar{\mathcal{O}}_L \mathcal{O}_L\rangle+ \frac{1}{2\sqrt{2}} \langle L_{-1}[O_f^{++} O_f^{+-}] \,[\bar{\mathcal{O}}_1 \bar{\mathcal{O}}_2]_P\,\bar{\mathcal{O}}_L \mathcal{O}_L\rangle\,,
  \end{aligned}
  \end{equation}
  and the first term in the second line can be rewritten as
  \begin{equation}
  \langle \mathcal{O}_1 \bar{\mathcal{O}}_1 \bar{\mathcal{O}}_L  \mathcal{O}_L\rangle=\mathcal{D}_{\frac{1}{2}}  \langle O_f^{++} O_f^{--} \bar{\mathcal{O}}_L  \mathcal{O}_L\rangle\,.
  \end{equation}
 Note that the mixed terms in the second line of \eqref{eq:primdesc} can be discarded for the purpose of our check, since they do not get contributions from the identity block. 
 
 The Virasoro algebra predicts that the $\bar{z}\to 1$ limit of a correlator with two primaries of dimension $h_f$ and two scalar primaries of dimension $h_g$ up to order $1/N^2$ is
 \begin{equation}
|1-z|^{4 h_g}\, \langle O_{h_f} \bar{O}_{h_f} \bar{O}_{h_g} O_{h_g} \rangle \to 1+\frac{1}{N} h_f h_g\,\mathcal{V}_1+\frac{1}{N^2} \left(h_f^2 h_g^2\, \mathcal{V}_2^{(2,2)}+(h_f^2 h_g +h_f h_g^2)\mathcal{V}_2^{(2,1)}+h_f h_g\,\mathcal{V}_2^{(1,1)} \right)\,,
 \end{equation}
where
\begin{equation}
\begin{aligned}
&\mathcal{V}_1 =-2\left(2+ \frac{1+z}{1-z}\,\log z\right)\quad , \quad \mathcal{V}_2^{(2,2)}=2\left(2+ \frac{1+z}{1-z}\log z\right)^2\,,\\
&\mathcal{V}_2^{(2,1)} = -2 \left(4+\frac{1+z}{1-z}\,\log z -  \frac{2\, z}{(1-z)^2}\,\log^2z \right)\,,\\
&\mathcal{V}_2^{(1,1)} = \frac{16}{3}-\frac{1}{3}\,\frac{1+z}{1-z}\,\log z + \frac{2\,z^2}{(1-z)^2}\,\log^2 z - \frac{4\,(1+z)}{(1-z)}\,\mathrm{Li}_2(1-z)\,.
\end{aligned}
\end{equation}
For primaries with non-vanishing R-charge, $j$,  the dimension $h$ to be used in the previous identities is actually the ``reduced" conformal dimension where one subtracts the Sugawara contribution: so we should take $h\to h- \frac{j^2}{N}$. In our computation we have $h_g=h_L=1$, while for $h_f$ we should consider several cases: for $[\mathcal{O}_1 \mathcal{O}_2]_P$, $h_f=2+\frac{\gamma_{[12]}}{N}-\frac{1}{N}$ with $\gamma_{[12]}$ the anomalous dimension of the non-BPS primary; for $[O_f^{++} O_f^{+-}] $, $h_f=1-\frac{1}{N}$; for $O_f^{++}$ and for $\mathcal{O}_2=O_f^{+-}$, $h_f=\frac{1}{2}-\frac{1}{4 N}$. Assembling all the pieces, one obtains a CFT prediction for the $\bar{z}\to 1$ limit of $\mathcal{C}^{(c)}_{[12]}$, as a function of the anomalous dimension $\gamma_{[12]}$:
\begin{equation}\label{eq:zbto1}
|1-z|^4\,\mathcal{C}^{(c)}_{[12]}\underset{\bar{z}\to 1}{\rightarrow} 2- 2\gamma_{[12]}+(10-\gamma_{[12]})\,\frac{1+z}{1-z}\,\log z + 3\,\frac{1+4z+z^2}{(1-z)^2}\,\log^2 z\,.
\end{equation}
The limit of the gravity result\footnote{Note that the correlators $\tilde{C}_{1,0}$ and $\tilde{C}_{0,1}$ contribute to the $\bar{z}\to 1$ limit with terms proportional to $\mathcal{V}_1$, and thus the values of the numerical coefficients, ($-\frac{5}{72}$, $\frac{1}{72}$), that multiply these correlators in \eqref{eq:Cc12} are crucial for the correct determination of the anomalous dimension.} for $\mathcal{C}^{(c)}_{[12]}$ has the same functional form:
\begin{equation}
|1-z|^4\,\mathcal{C}^{(c)}_{[12]}\underset{\bar{z}\to 1}{\rightarrow} \,\frac{16}{3}+\frac{35}{3}\frac{1+z}{1-z}\,\log z + 3\,\frac{1+4z+z^2}{(1-z)^2}\,\log^2 z\,,
\end{equation}
and the two expressions can be matched by taking
\begin{equation}\label{eq:andimpred}
    \gamma_{[12]}= -\frac{5}{3}\,.
\end{equation}
As a further consistency check, one can verify that the same prediction for $\gamma_{[12]}$ comes by looking at the limit $z\to 1$ with fixed $\bar z$. The computation is similar but simpler than the previous one, because all the states involved in the correlator $\mathcal{C}^{(c)}_{[12]}$ are primaries with respect to the right-moving algebra. The relevant right-moving dimensions are: $\bar{h}_g=1$ for $\mathcal{O}_L$, $\bar{h}_f=1+\frac{\gamma_{[12]}}{N}$ for $[\mathcal{O}_1 \mathcal{O}_2]_P$, $\bar{h}_f=1$ for $L_{-1}(O_f^{++} O_f^{+-})$ and $\bar{h}_f=\frac{1}{2}-\frac{1}{4N}$ for $\mathcal{O}_1$ and $\mathcal{O}_2$. This leads to 
\begin{equation}\label{eq:zto1}
|1-z|^4\,\mathcal{C}^{(c)}_{[12]}\underset{z\to 1}{\rightarrow} -2(1+\,\gamma_{[12]})+\,(2-\gamma_{[12]})\,\frac{1+\bar z}{1-\bar z}\,\log \bar z+\frac{1+4\bar z+\bar z^2}{(1-\bar z)^2}\,\log^2 \bar z\,,
\end{equation}
which agrees with the supergravity result
\begin{equation}
|1-z|^4\,\mathcal{C}^{(c)}_{[12]}\underset{z\to 1}{\rightarrow} \frac{4}{3}+\frac{11}{3} \,\frac{1+\bar z}{1-\bar z}\,\log \bar z+\frac{1+4\bar z+\bar z^2}{(1-\bar z)^2}\,\log^2 \bar z\,,
\end{equation}
if we use again the value \eqref{eq:andimpred} for $\gamma_{[12]}$.

The ones above are already non-trivial checks of the gravity result, but the strongest test comes from verifying the prediction \eqref{eq:andimpred} for the anomalous dimension. In the following Section we will extract the value of $\gamma_{[12]}$ from the OPE of a known 4-point correlator with only single-particle operators and we will compare with \eqref{eq:andimpred}.

\subsection{Derivation of the anomalous dimension of $[\mathcal{O}_1 \mathcal{O}_2]_P$}
The double-particle operator $[\mathcal{O}_1 \mathcal{O}_2]_P$ is exchanged in the $z\to 0$ OPE channel of the correlator
\begin{equation}
    \langle O^{++}_{f_1}(0) O^{--}_{f_2}(\infty) O^{-+}_{f_3}(1) O^{+-}_{f_4}(z,\bar{z})\rangle \,,
\end{equation}
where it is useful to keep the flavour indices, $f_i$, generic, as we will see. 
We should ask if this is the only primary exchanged with bare dimension $(h,\bar{h})=(2,1)$. At the gravity point, where we want to compute $\gamma_{[12]}$, the CFT acquires an $SO(N_f)$ flavour symmetry, which implies that the 4-point correlator has the form
\begin{equation}\label{eq:4point}
 \langle O^{++}_{f_1}(0) O^{--}_{f_2}(\infty) O^{-+}_{f_3}(1) O^{+-}_{f_4}(z,\bar{z})\rangle=\frac{1}{|1-z|^2}\left(\delta_{f_1 f_2}\delta_{f_3 f_4}\,\mathcal{G}_{12}+\delta_{f_1 f_3}\delta_{f_2 f_4}\,\mathcal{G}_{13}+\delta_{f_1 f_4}\delta_{f_2 f_3}\,\mathcal{G}_{14} \right)\,,
\end{equation}
and that the primaries must sit into irreducible representations of $SO(N_f)$. For the case of $[\mathcal{O}_1 \mathcal{O}_2]_P$, which is in the symmetrized tensor product of two fundamental representations of $SO(N_f)$, the relevant irreducible representations are the symmetric traceless and the identity representation:
\begin{equation}\label{eq:symandid}
   [\mathcal{O}_1 \mathcal{O}_2]_P =\sqrt{1-\frac{1}{N_f}}\, \mathcal{O}_{\mathrm{sym}} + \sqrt{\frac{1}{N_f}}\,  \mathcal{O}_{\mathrm{id}}\,,
\end{equation}
where $\mathcal{O}_{\mathrm{sym}}$ and $\mathcal{O}_{\mathrm{id}}$ denote the two normalised primaries in the respective irreps. Note that $[\mathcal{O}_1 \mathcal{O}_2]_P$ being anti-symmetric with respect to the exchange of the right-moving R-symmetry indices (see \eqref{eq:primdef}), is already in the $(1,0)$ irrep of $SU(2)_L \times SU(2)_R$. The primaries $\mathcal{O}_{\mathrm{sym}}$ and $\mathcal{O}_{\mathrm{id}}$ are the only two that appear in the OPE of \eqref{eq:4point} at dimension $(h,\bar{h})=(2,1)$ and thus their anomalous dimension can be read off directly from the projections of \eqref{eq:4point} to the respective flavour representations:
\begin{equation}
    \mathcal{C}_\mathrm{sym} =\frac{1}{|1-z|^2}\,(\mathcal{G}_{12}+\mathcal{G}_{13})\quad,\quad \mathcal{C}_\mathrm{id} = \frac{1}{|1-z|^2}\,(\mathcal{G}_{12}+\mathcal{G}_{13}+N_f\,\mathcal{G}_{14})\,.
\end{equation}
The supergravity correlator \eqref{eq:4point} was computed holographically in \cite{Giusto:2018ovt} and then reproduced with a different method in \cite{Rastelli:2019gtj}. It is given by:
\begin{equation}
    \mathcal{G}_{12}=1-\frac{1}{N}+\frac{z}{N}\,\bar{D}_{1122}\quad,\quad \mathcal{G}_{13}=\frac{|1-z|^2}{\bar{z}\,N}\,\bar{D}_{2112}\quad,\quad \mathcal{G}_{14}=\frac{z\,|1-z|^2}{N}\,\bar{D}_{1212}\,.
\end{equation}
As there is no mixing with other operators, the anomalous dimensions can be obtained from this correlator in a straightforward way: one expands $\mathcal{C}_\mathrm{sym}$ and $\mathcal{C}_\mathrm{id}$ up to order $z^1\,\bar{z}^0$ and subtracts the conformal block corresponding to the primary exchanged at order $z^0\,\bar{z}^0$; in what remains at order $z^1\,\bar{z}^0$, the ratio between the term proportional to $\log|z|^2$ and the algebraic term gives the anomalous dimensions, $\gamma_\mathrm{sym}$ and $\gamma_\mathrm{id}$. The result of this calculation is
\begin{equation}
    \gamma_\mathrm{sym}=-\frac{2}{3}\quad,\quad \gamma_\mathrm{id}= -\frac{2}{3}-N_f\,.
\end{equation}

The dimension $\gamma_{[12]}$ that appears in \eqref{eq:zbto1} and \eqref{eq:zto1} is the combination of $\gamma_\mathrm{sym}$ and $\gamma_\mathrm{id}$ determined by the coefficients in \eqref{eq:symandid}:
\begin{equation}
    \gamma_{[12]}= \left(1-\frac{1}{N_f}\right)\,\gamma_\mathrm{sym}+\frac{1}{N_f}\, \gamma_\mathrm{id}=-\frac{5}{3}\,.
\end{equation}
The fact that this value matches the gravity prediction \eqref{eq:andimpred} is the main result of our analysis. Note also that it is consistent that the $N_f$ dependence of $\gamma_\mathrm{id}$ cancels in the combination $\gamma_{[12]}$, because the gravity solution from which we have extracted our correlator is trivial in the 4D compact space and does not distinguish $T^4$ from $K3$. 

\section{Concluding remarks}
\label{sec:concl}

We have provided evidence that the light limit ($\alpha_i \to 0$) of a HHLL supergravity correlator computes the tree-level connected part of correlators between two $n$-particle and two single-particle operators at order $c^{-n}$ in the large $c$ expansion, even when the multi-particles are non-supersymmetric. To reconstruct the full correlator at order $c^{-n}$ one should add the disconnected loop contributions, which cannot be computed in classical supergravity.

The most non-trivial test we could provide for this conjecture consists in the agreement between the anomalous dimension of a non-BPS double-particle operator computed in two independent ways: on one hand, by analyzing the identity block of the 4-point function containing the double-particles, derived from the light limit of the HHLL correlator; and, on the other hand, by studying the exchange of the double-particle in a 4-point correlator with only single-particle operators. It is interesting to compare this anomalous dimension with a quantity superficially similar to it, computed in \cite{Ganchev:2023sth}. The quantity denoted as $\delta \omega_2$ in eq. (5.8) of that paper, evaluated for $n_1=1$, $n_2=0$, represents the interaction energy between a particle of type $\mathcal{O}_2$ and $N_1$ particles of type $\mathcal{O}_1$ and is derived by studying the angular momenta and mass of the supergravity solution dual to the coherent state $O_H$ in \eqref{eq:OHO1O2}, in the limit $\alpha_i\to 0$. Hence, the result in eq. (5.8) can be trusted in the regime in which $N_1\sim N\gg 1$ and $N_1/N\ll 1$. Nevertheless, if one extrapolates that result to the regime appropriate to describe the double-particle $[\mathcal{O}_1 \mathcal{O}_2]$, i.e. if one takes $N_1=1$, one obtains $\delta\omega_2 = -\frac{1}{N} \frac{5}{3}$. To compare with \eqref{eq:andimpred}, we should recall that $[\mathcal{O}_1 \mathcal{O}_2]$ is the linear combination, \eqref{eq:primdef}, of the primary $[\mathcal{O}_1 \mathcal{O}_2]_P$, with anomalous dimension $\gamma_{[12]}$, and a descendant with vanishing anomalous dimension, and thus the effective left-moving anomalous dimension of $[\mathcal{O}_1 \mathcal{O}_2]$ is $-\frac{5}{6}$ and the total left plus right dimension is $-\frac{5}{3}$. This agrees with the $\delta\omega_2$ derived above, which represents the correction to the energy and, hence, to the total (left plus right) dimension. Thus, at least in this example, taking the light limit of the anomalous dimension of the heavy state does lead to the correct result for the anomalous dimension of a $n$-particle operator with finite $n$ in the large $N$ limit. This agrees with the fact that the light limit of the HHLL supergravity correlator correctly captures the tree-level connected part of the correlators with $n$-particle operators with finite $n$. 

If we trust this conjecture, we can apply our supergravity method to derive the correlators with more general choices for the non-BPS multi-particle operators, as far as we are able to construct the associated geometries perturbatively in $\alpha$. As we have mentioned at the start of Section~\ref{sec:nonBPScorr}, we have already computed the connected part of the correlator
\begin{equation}\label{eq:LLbarcor}
    \langle [L_{-1}\tilde{L}_{-1} O^{++}_f]^2 \,[L_{-1}\tilde{L}_{-1} O^{--}_f]^2\,\bar{\mathcal{O}}_L \,\mathcal{O}_L \rangle\,,  
\end{equation}
with $\mathcal{O}_L$ the same operator as in \eqref{eq:OLdef}. We provide some details on how to extract this correlator from gravity in Appendix~\ref{sec:appA}. The correlator \eqref{eq:LLbarcor} has the same form as in \eqref{eq:C12conn}, and we provide the corresponding rational functions, $R_i$, in the ancillary file. One could extract the anomalous dimension\footnote{This double-particle operator is again the linear combination of some primaries and some descendants, so the anomalous dimension following from \eqref{eq:LLbarcor} will be the corresponding combination of the dimensions of the primaries.} of $[L_{-1}\tilde{L}_{-1} O^{++}_f]^2$ with the same method described above. It would be interesting to compare this prediction with the anomalous dimension computed with the method of \cite{Aprile:2021mvq,Aprile:2026yit}.

As a final note, the knowledge of holographic correlators with external non-BPS multi-particle operators allows the computation of a new class of non-protected CFT data at strong coupling, namely 3-point functions containing two or three non-BPS operators. 

\section*{Acknowledgements}
We are very grateful to Rodolfo Russo for many enlightening discussions on holographic correlators, for a careful reading of the manuscript, and for valuable comments and suggestions. It is also a pleasure to thank Francesco Aprile for useful comments on the anomalous dimensions. M.G. is supported by the Swiss National Science Foundation Grant No.~218510 .

\appendix

\section{Details on the \texorpdfstring{$[L_{-1}\tilde{L}_{-1} O^{++}_f]^2$}{LLbarO} correlator}
\label{sec:appA}
For completeness, we include in this appendix more details on how the correlator \eqref{eq:LLbarcor} was evaluated. We start from the $\alpha$ class of solutions in \cite{Ganchev:2021ewa}. Following their conventions, we denote by $N_1$ the number of single-particle constituents that appear in the heavy operator, while $n$ and $m$ denote two integers that identify a geometry in that class of solutions. Here we are interested in the $n=0$, $m=1$ case.  Once the supergravity solution is known up to order $\tilde\alpha^4$, we find the correlator $\mathcal C_H^{\rm sugra}(z,\bar z;\tilde\alpha)$ following the same steps as above. The only non-trivial step is finding the relation between the gravitational parameter $\tilde\alpha^2$ (which in \cite{Ganchev:2021ewa} was denoted by $\alpha^2$) and the CFT parameter $\alpha^2 \equiv N_1/N$, which is necessary to identify the CFT correlator. We expect a relation of the following form
\begin{equation}
    \tilde\alpha^2=4\alpha^2\left(1 + c_\alpha\alpha^2+O(\alpha^4)\right)\,.
    \label{def:c_alpha}
\end{equation}
In \cite{Ganchev:2021ewa} the value of $c_\alpha$ was derived for $m=1$ and $n\neq 0$ (Eq. (4.14)). Once this relation was known, it was possible to evaluate the frequency shift $\delta\omega$ and the total energy up to order $\alpha^4$. The former can be evaluated after performing a large gauge transformation $\varphi_1 \to \varphi_1 -\delta\omega \, \tau$. Then, the constraint $j=N_1/2$, together with \eqref{def:c_alpha}, fixes $\delta\omega$ to
\begin{equation}
\delta\omega = -c_\omega \alpha^2 + O(\alpha^4)\,,
\label{deltaOmega}
\end{equation}
where $c_\omega$ is a rational number that depends on $n$. On the other hand, the total energy can be evaluated using holographic prescriptions that relate $(h,\bar h)$ to the metric and the gauge fields. By doing so, one finds that there is a correction to the total energy with respect to the free result, $3\,N_1$, which was expected to be related to the frequency shift $\delta\omega$. Indeed, this comparison was used as a non-trivial check in the $n\neq0$ case, leading to the relation
\begin{equation}
h+\bar h = N_1 \left(3+\frac{\delta\omega}{2}+O(\alpha^4)\right)\,.
\label{h_plus_hbar}
\end{equation}
In the $n=0$ case, the known formula for $c_\alpha$ in \eqref{def:c_alpha} degenerates, so we cannot use that equation. However, we expect that equation \eqref{h_plus_hbar} holds also in this case. Thus, we can reverse the argument: we assume that the quantity $\delta\omega$ appearing in \eqref{deltaOmega} and \eqref{h_plus_hbar} is the same, and we use these relations to fix $c_\alpha$. For $n=0$, we find
\begin{equation}
    \delta\omega = - \frac{24}{5}\alpha^2 + O(\alpha^4) \quad \implies\quad c_\omega = \frac{24}{5}\,.
	\label{deltaOmega_2}
\end{equation}
Note that this result coincides with the one found in \cite{Ganchev:2021ewa}, since at this order the parameter $c_\alpha$ does not enter the equations. Now consider the sum $h+\bar h$, which can be evaluated using holographic techniques in terms of the metric and the gauge fields. In our solution ($m=1$, $n=0$), the relevant functions read
\begin{equation}
    \begin{aligned}
        g_{\tau\tau}^{(2)} & = -\frac 1 4 + \frac{7}{16}\tilde\alpha^2 - \frac{1027}{3200} \tilde\alpha^4 +O(\tilde\alpha^6)\,\\
        g_{\sigma\sigma}^{(2)} & = -\frac{3}{4}+ \frac{17}{16}\tilde\alpha^2 - \frac{73}{640}\tilde\alpha^4 +O(\tilde\alpha^6)\,\\
        g_{\tau\sigma}^{(2)} & = O(\tilde\alpha^6)
    \end{aligned}
\end{equation}
and\footnote{Note that our gauge fields differ from those given in \cite[eq.~(3.22)]{Ganchev:2021ewa}. Indeed, the constraint $A_\tau^{\varphi_2}(0)=0$ imposed in that solution does not hold for $n=0$, so we had to perform a gauge transformation in this case.}
\begin{equation}
    \begin{aligned}
        A_\tau^{(0)\pm}&=\frac{29}{20}\tilde\alpha^2 + \delta\omega + O(\tilde\alpha^6)\,\\
        A_\sigma^{(0)\pm}&=\pm\frac{1}{4}\tilde\alpha^2\left(1 -\frac{29}{600}\tilde\alpha^2\right) +O(\tilde\alpha^6)\,.
    \end{aligned}
\end{equation}
These quantities yield
\begin{equation}
    h + \bar h = \frac{N}{4}\left(3\tilde\alpha^2 + \frac{23}{40}\tilde\alpha^4 + \frac{17}{10}\tilde\alpha^2\delta\omega+\frac{1}{2}\delta\omega^2+O(\tilde\alpha^6)\right)\,.
\end{equation}
By comparing this result with \eqref{h_plus_hbar}, we find
\begin{equation}
    4\alpha^2\left(3 + \frac{\delta\omega}{2}\right)  = 3 \tilde\alpha^2+ \frac{23}{40}\tilde\alpha^4 + \frac{17}{10}\tilde\alpha^2\delta\omega+\frac{1}{2}\delta\omega^2 + O(\alpha^6)\,.
\end{equation}
Using \eqref{def:c_alpha} and \eqref{deltaOmega_2} the terms at order $\alpha^2$ cancel, leaving an equation for $c_\alpha$ at order $\alpha^4$, which gives
\begin{equation}
    c_\alpha = \frac{29}{150}\,.
\end{equation}
Note that this result is different from the naive value obtained by taking the limit $n\to0$ in (4.14), which would be $179/150$. Finally, we can use this result to extract the correlator \eqref{eq:LLbarcor}, which we denote by $\mathcal C^{(c)}_{2}$. In this case the geometry depends on a single parameter, $\tilde\alpha$, and thus the equation \eqref{eq:Cnm} becomes
\begin{equation}
\mathcal C_H^{\rm sugra}(z,\bar z;\tilde \alpha) = |1-z|^{-4} + \sum_{k=1}\frac{\tilde\alpha^{2k}}{k!} N^k \tilde C_{k}(z,\bar z)\,.
\end{equation}
Using \eqref{eq:CHsugrasmallalpha} together with  \eqref{def:c_alpha} to rewrite $\tilde\alpha$ in terms of $\alpha$, we can extract $\mathcal C^{(c)}_{2}$, finding
\begin{equation}
\mathcal C^{(c)}_{2} = 16\left(\tilde C_{2} + \frac{29}{300}\, \frac{\tilde C_{1}}{N}\right)\,.
\end{equation}

\bibliographystyle{utphys}

\begin{thebibliography}{10}

\bibitem{Maldacena:1997re}
J.~M. Maldacena, ``{The Large N limit of superconformal field theories and
  supergravity},'' \href{http://dx.doi.org/10.4310/ATMP.1998.v2.n2.a1}{{\em
  Adv. Theor. Math. Phys.} {\bfseries 2} (1998) 231--252},
  \href{http://arxiv.org/abs/hep-th/9711200}{{\ttfamily arXiv:hep-th/9711200}}.

\bibitem{Witten:1998qj}
E.~Witten, ``{Anti-de Sitter space and holography},''
  \href{http://dx.doi.org/10.4310/ATMP.1998.v2.n2.a2}{{\em Adv. Theor. Math.
  Phys.} {\bfseries 2} (1998) 253--291},
  \href{http://arxiv.org/abs/hep-th/9802150}{{\ttfamily arXiv:hep-th/9802150}}.

\bibitem{Aprile:2017xsp}
F.~Aprile, J.~M. Drummond, P.~Heslop, and H.~Paul, ``{Unmixing Supergravity},''
  \href{http://dx.doi.org/10.1007/JHEP02(2018)133}{{\em JHEP} {\bfseries 02}
  (2018) 133}, \href{http://arxiv.org/abs/1706.08456}{{\ttfamily
  arXiv:1706.08456 [hep-th]}}.

\bibitem{Aprile:2021mvq}
F.~Aprile and M.~Santagata, ``{Two particle spectrum of tensor multiplets
  coupled to AdS3{\texttimes}S3 gravity},''
  \href{http://dx.doi.org/10.1103/PhysRevD.104.126022}{{\em Phys. Rev. D}
  {\bfseries 104} no.~12, (2021) 126022},
  \href{http://arxiv.org/abs/2104.00036}{{\ttfamily arXiv:2104.00036
  [hep-th]}}.

\bibitem{Aprile:2026yit}
F.~Aprile, H.~Paul, and M.~Santagata, ``{Quantum gravity on
  AdS$_{3}${\texttimes}S$^{3}$ from CFT: bootstrapping n = 21},''
  \href{http://dx.doi.org/10.1007/JHEP08(2026)003}{{\em JHEP} {\bfseries 08}
  (2026) 003}, \href{http://arxiv.org/abs/2602.11254}{{\ttfamily
  arXiv:2602.11254 [hep-th]}}.

\bibitem{Aprile:2026uxe}
F.~Aprile, S.~Giusto, R.~Russo, and J.~Vilas~Boas, ``{Multi-particle
  correlators with higher KK modes I: a bootstrap approach},''
  \href{http://arxiv.org/abs/2601.16085}{{\ttfamily arXiv:2601.16085
  [hep-th]}}.

\bibitem{Goncalves:2019znr}
V.~Gon\c{c}alves, R.~Pereira, and X.~Zhou, ``{$20'$ Five-Point Function from
  $AdS_5\times S^5$ Supergravity},''
  \href{http://dx.doi.org/10.1007/JHEP10(2019)247}{{\em JHEP} {\bfseries 10}
  (2019) 247}, \href{http://arxiv.org/abs/1906.05305}{{\ttfamily
  arXiv:1906.05305 [hep-th]}}.

\bibitem{Goncalves:2023oyx}
V.~Gon\c{c}alves, C.~Meneghelli, R.~Pereira, J.~Vilas~Boas, and X.~Zhou,
  ``{Kaluza-Klein five-point functions from AdS$_{5}$\texttimes{}S$^{5}$
  supergravity},'' \href{http://dx.doi.org/10.1007/JHEP08(2023)067}{{\em JHEP}
  {\bfseries 08} (2023) 067}, \href{http://arxiv.org/abs/2302.01896}{{\ttfamily
  arXiv:2302.01896 [hep-th]}}.

\bibitem{Ceplak:2021wzz}
N.~Ceplak, S.~Giusto, M.~R.~R. Hughes, and R.~Russo, ``{Holographic correlators
  with multi-particle states},''
  \href{http://dx.doi.org/10.1007/JHEP09(2021)204}{{\em JHEP} {\bfseries 09}
  (2021) 204}, \href{http://arxiv.org/abs/2105.04670}{{\ttfamily
  arXiv:2105.04670 [hep-th]}}.

\bibitem{Aprile:2024lwy}
F.~Aprile, S.~Giusto, and R.~Russo, ``{Holographic correlators with BPS bound
  states in $\mathcal{N} = 4$ SYM},''
  \href{http://dx.doi.org/10.1103/PhysRevLett.134.091602}{{\em Phys. Rev.
  Lett.} {\bfseries 134} no.~9, (2025) 091602},
  \href{http://arxiv.org/abs/2409.12911}{{\ttfamily arXiv:2409.12911
  [hep-th]}}.

\bibitem{Aprile:2025hlt}
F.~Aprile, S.~Giusto, and R.~Russo, ``{Four-point correlators with BPS bound
  states in AdS$_{3}$ and AdS$_{5}$},''
  \href{http://dx.doi.org/10.1007/JHEP08(2025)193}{{\em JHEP} {\bfseries 08}
  (2025) 193}, \href{http://arxiv.org/abs/2503.02855}{{\ttfamily
  arXiv:2503.02855 [hep-th]}}.

\bibitem{Shigemori:2020yuo}
M.~Shigemori, ``{Superstrata},''
  \href{http://dx.doi.org/10.1007/s10714-020-02698-8}{{\em Gen. Rel. Grav.}
  {\bfseries 52} no.~5, (2020) 51},
  \href{http://arxiv.org/abs/2002.01592}{{\ttfamily arXiv:2002.01592
  [hep-th]}}.

\bibitem{Bena:2022rna}
I.~Bena, E.~J. Martinec, S.~D. Mathur, and N.~P. Warner, ``{Fuzzballs and
  Microstate Geometries: Black-Hole Structure in String Theory},''
  \href{http://arxiv.org/abs/2204.13113}{{\ttfamily arXiv:2204.13113
  [hep-th]}}.

\bibitem{Lunin:2001jy}
O.~Lunin and S.~D. Mathur, ``{AdS / CFT duality and the black hole information
  paradox},'' \href{http://dx.doi.org/10.1016/S0550-3213(01)00620-4}{{\em Nucl.
  Phys. B} {\bfseries 623} (2002) 342--394},
  \href{http://arxiv.org/abs/hep-th/0109154}{{\ttfamily arXiv:hep-th/0109154}}.

\bibitem{Bena:2015bea}
I.~Bena, S.~Giusto, R.~Russo, M.~Shigemori, and N.~P. Warner, ``{Habemus
  Superstratum! A constructive proof of the existence of superstrata},''
  \href{http://dx.doi.org/10.1007/JHEP05(2015)110}{{\em JHEP} {\bfseries 05}
  (2015) 110}, \href{http://arxiv.org/abs/1503.01463}{{\ttfamily
  arXiv:1503.01463 [hep-th]}}.

\bibitem{Bena:2016ypk}
I.~Bena, S.~Giusto, E.~J. Martinec, R.~Russo, M.~Shigemori, D.~Turton, and
  N.~P. Warner, ``{Smooth horizonless geometries deep inside the black-hole
  regime},'' \href{http://dx.doi.org/10.1103/PhysRevLett.117.201601}{{\em Phys.
  Rev. Lett.} {\bfseries 117} no.~20, (2016) 201601},
  \href{http://arxiv.org/abs/1607.03908}{{\ttfamily arXiv:1607.03908
  [hep-th]}}.

\bibitem{Bena:2017xbt}
I.~Bena, S.~Giusto, E.~J. Martinec, R.~Russo, M.~Shigemori, D.~Turton, and
  N.~P. Warner, ``{Asymptotically-flat supergravity solutions deep inside the
  black-hole regime},'' \href{http://dx.doi.org/10.1007/JHEP02(2018)014}{{\em
  JHEP} {\bfseries 02} (2018) 014},
  \href{http://arxiv.org/abs/1711.10474}{{\ttfamily arXiv:1711.10474
  [hep-th]}}.

\bibitem{Heidmann:2019zws}
P.~Heidmann and N.~P. Warner, ``{Superstratum Symbiosis},''
  \href{http://dx.doi.org/10.1007/JHEP09(2019)059}{{\em JHEP} {\bfseries 09}
  (2019) 059}, \href{http://arxiv.org/abs/1903.07631}{{\ttfamily
  arXiv:1903.07631 [hep-th]}}.

\bibitem{Heidmann:2019xrd}
P.~Heidmann, D.~R. Mayerson, R.~Walker, and N.~P. Warner, ``{Holomorphic Waves
  of Black Hole Microstructure},''
  \href{http://dx.doi.org/10.1007/JHEP02(2020)192}{{\em JHEP} {\bfseries 02}
  (2020) 192}, \href{http://arxiv.org/abs/1910.10714}{{\ttfamily
  arXiv:1910.10714 [hep-th]}}.

\bibitem{Houppe:2020oqp}
A.~Houppe and N.~P. Warner, ``{Supersymmetry and superstrata in three
  dimensions},'' \href{http://dx.doi.org/10.1007/JHEP08(2021)133}{{\em JHEP}
  {\bfseries 08} (2021) 133}, \href{http://arxiv.org/abs/2012.07850}{{\ttfamily
  arXiv:2012.07850 [hep-th]}}.

\bibitem{Ganchev:2021iwy}
B.~Ganchev, A.~Houppe, and N.~P. Warner, ``{New superstrata from
  three-dimensional supergravity},''
  \href{http://dx.doi.org/10.1007/JHEP04(2022)065}{{\em JHEP} {\bfseries 04}
  (2022) 065}, \href{http://arxiv.org/abs/2110.02961}{{\ttfamily
  arXiv:2110.02961 [hep-th]}}.

\bibitem{Ganchev:2022exf}
B.~Ganchev, A.~Houppe, and N.~P. Warner, ``{Elliptical and purely NS
  superstrata},'' \href{http://dx.doi.org/10.1007/JHEP09(2022)067}{{\em JHEP}
  {\bfseries 09} (2022) 067}, \href{http://arxiv.org/abs/2207.04060}{{\ttfamily
  arXiv:2207.04060 [hep-th]}}.

\bibitem{Lin:2004nb}
H.~Lin, O.~Lunin, and J.~M. Maldacena, ``{Bubbling AdS space and 1/2 BPS
  geometries},'' \href{http://dx.doi.org/10.1088/1126-6708/2004/10/025}{{\em
  JHEP} {\bfseries 10} (2004) 025},
  \href{http://arxiv.org/abs/hep-th/0409174}{{\ttfamily arXiv:hep-th/0409174}}.

\bibitem{Liu:2007xj}
J.~T. Liu, H.~Lu, C.~N. Pope, and J.~F. Vazquez-Poritz, ``{Bubbling AdS black
  holes},'' \href{http://dx.doi.org/10.1088/1126-6708/2007/10/030}{{\em JHEP}
  {\bfseries 10} (2007) 030},
  \href{http://arxiv.org/abs/hep-th/0703184}{{\ttfamily arXiv:hep-th/0703184}}.

\bibitem{Giusto:2024trt}
S.~Giusto and A.~Rosso, ``{The geometry of large charge multi-traces in
  $\mathcal{N}=4$ SYM},'' \href{http://arxiv.org/abs/2401.01254}{{\ttfamily
  arXiv:2401.01254 [hep-th]}}.

\bibitem{Ganchev:2021ewa}
B.~Ganchev, S.~Giusto, A.~Houppe, and R.~Russo, ``{$\hbox {AdS}_3$ holography
  for non-BPS geometries},''
  \href{http://dx.doi.org/10.1140/epjc/s10052-022-10133-2}{{\em Eur. Phys. J.
  C} {\bfseries 82} no.~3, (2022) 217},
  \href{http://arxiv.org/abs/2112.03287}{{\ttfamily arXiv:2112.03287
  [hep-th]}}.

\bibitem{Ganchev:2023sth}
B.~Ganchev, S.~Giusto, A.~Houppe, R.~Russo, and N.~P. Warner,
  ``{Microstrata},'' \href{http://dx.doi.org/10.1007/JHEP10(2023)163}{{\em
  JHEP} {\bfseries 10} (2023) 163},
  \href{http://arxiv.org/abs/2307.13021}{{\ttfamily arXiv:2307.13021
  [hep-th]}}.

\bibitem{Galliani:2017jlg}
A.~Galliani, S.~Giusto, and R.~Russo, ``{Holographic 4-point correlators with
  heavy states},'' \href{http://dx.doi.org/10.1007/JHEP10(2017)040}{{\em JHEP}
  {\bfseries 10} (2017) 040}, \href{http://arxiv.org/abs/1705.09250}{{\ttfamily
  arXiv:1705.09250 [hep-th]}}.

\bibitem{Bombini:2017sge}
A.~Bombini, A.~Galliani, S.~Giusto, E.~Moscato, and R.~Russo, ``{Unitary
  4-point correlators from classical geometries},''
  \href{http://dx.doi.org/10.1140/epjc/s10052-017-5492-3}{{\em Eur. Phys. J. C}
  {\bfseries 78} no.~1, (2018) 8},
  \href{http://arxiv.org/abs/1710.06820}{{\ttfamily arXiv:1710.06820
  [hep-th]}}.

\bibitem{Giusto:2018ovt}
S.~Giusto, R.~Russo, and C.~Wen, ``{Holographic correlators in AdS$_{3}$},''
  \href{http://dx.doi.org/10.1007/JHEP03(2019)096}{{\em JHEP} {\bfseries 03}
  (2019) 096}, \href{http://arxiv.org/abs/1812.06479}{{\ttfamily
  arXiv:1812.06479 [hep-th]}}.

\bibitem{Giusto:2019pxc}
S.~Giusto, R.~Russo, A.~Tyukov, and C.~Wen, ``{Holographic correlators in
  AdS$_3$ without Witten diagrams},''
  \href{http://dx.doi.org/10.1007/JHEP09(2019)030}{{\em JHEP} {\bfseries 09}
  (2019) 030}, \href{http://arxiv.org/abs/1905.12314}{{\ttfamily
  arXiv:1905.12314 [hep-th]}}.

\bibitem{Turton:2024afd}
D.~Turton and A.~Tyukov, ``{Four-point correlators in N=4 SYM from AdS$_5$
  bubbling geometries},'' \href{http://arxiv.org/abs/2408.16834}{{\ttfamily
  arXiv:2408.16834 [hep-th]}}.

\bibitem{Turton:2025cnn}
D.~Turton and A.~Tyukov, ``{Holographic correlators from multi-mode AdS$_{5}$
  bubbling geometries},'' \href{http://dx.doi.org/10.1007/JHEP05(2026)102}{{\em
  JHEP} {\bfseries 05} (2026) 102},
  \href{http://arxiv.org/abs/2512.19392}{{\ttfamily arXiv:2512.19392
  [hep-th]}}.

\bibitem{Maldacena:1998bw}
J.~M. Maldacena and A.~Strominger, ``{AdS(3) black holes and a stringy
  exclusion principle},''
  \href{http://dx.doi.org/10.1088/1126-6708/1998/12/005}{{\em JHEP} {\bfseries
  12} (1998) 005}, \href{http://arxiv.org/abs/hep-th/9804085}{{\ttfamily
  arXiv:hep-th/9804085}}.

\bibitem{Ganchev:2021pgs}
B.~Ganchev, A.~Houppe, and N.~P. Warner, ``{Q-balls meet fuzzballs: non-BPS
  microstate geometries},''
  \href{http://dx.doi.org/10.1007/JHEP11(2021)028}{{\em JHEP} {\bfseries 11}
  (2021) 028}, \href{http://arxiv.org/abs/2107.09677}{{\ttfamily
  arXiv:2107.09677 [hep-th]}}.

\bibitem{Mayerson:2020tcl}
D.~R. Mayerson, R.~A. Walker, and N.~P. Warner, ``{Microstate Geometries from
  Gauged Supergravity in Three Dimensions},''
  \href{http://dx.doi.org/10.1007/JHEP10(2020)030}{{\em JHEP} {\bfseries 10}
  (2020) 030}, \href{http://arxiv.org/abs/2004.13031}{{\ttfamily
  arXiv:2004.13031 [hep-th]}}.

\bibitem{Isaev:2003tk}
A.~P. Isaev, ``{Multiloop Feynman integrals and conformal quantum mechanics},''
  \href{http://dx.doi.org/10.1016/S0550-3213(03)00393-6}{{\em Nucl. Phys. B}
  {\bfseries 662} (2003) 461--475},
  \href{http://arxiv.org/abs/hep-th/0303056}{{\ttfamily arXiv:hep-th/0303056}}.

\bibitem{Fitzpatrick:2015qma}
A.~L. Fitzpatrick, J.~Kaplan, M.~T. Walters, and J.~Wang, ``{Eikonalization of
  Conformal Blocks},'' \href{http://dx.doi.org/10.1007/JHEP09(2015)019}{{\em
  JHEP} {\bfseries 09} (2015) 019},
  \href{http://arxiv.org/abs/1504.01737}{{\ttfamily arXiv:1504.01737
  [hep-th]}}.

\bibitem{Rastelli:2019gtj}
L.~Rastelli, K.~Roumpedakis, and X.~Zhou, ``{$\mathbf{AdS_3\times S^3}$
  Tree-Level Correlators: Hidden Six-Dimensional Conformal Symmetry},''
  \href{http://dx.doi.org/10.1007/JHEP10(2019)140}{{\em JHEP} {\bfseries 10}
  (2019) 140}, \href{http://arxiv.org/abs/1905.11983}{{\ttfamily
  arXiv:1905.11983 [hep-th]}}.

\end{thebibliography}

\providecommand{\href}[2]{#2}\begingroup\raggedright\endgroup

\end{document}